# Replacement policy via integration of dynamic programming and simulated annealing algorithm for systems with dependent components

Hamed Badakhsh[1], M.Sc.

Email: badakhsh_hamed@ie.sharif.edu

Mohammad Pirhooshyaran[1], M.Sc.

Email: pirhooshyaran_m@ie.sharif.edu

Abdolhamid Eshragh-Jahromi[1, 2] Ph.D.

P.O. Box 11155-9414 Azadi Ave., Tehran 1458889694 Iran

Tel: (+9821) 66165705, Fax: (+9821) 66022702 Email: eshragh@Sharif.edu

## Abstract

In a dependent multi-component system, increasing the deterioration of a part leads to the increased deterioration rate of other parts as well. In these systems, a deterioration limit is usually pre-determined for each part and the considered part is replaced while reaching this limit. In this paper, replacement conditions of these parts were examined according to the replacement times in the past. Using dynamic programming, it has been shown that for each deterioration rate of part 1, there is a deterioration limit for part 2, which can lead either part 2 or both parts should be replaced. The only available system data are the replacement time of the parts in the past according to the replacement policy at the time of reaching deterioration limit. Therefore, simulated annealing optimization method was used for estimating deterioration rates. Finally, two examples were presented for comparing the proposed method with the special limit replacement method, which showed the significance superiority of the former.

[1] Department of Industrial Engineering, Sharif University of Technology

[2] Corresponding Author

## 1. Introduction

The parts of a system may be dependent or independent. If they are independent, they can be separately planned for; however, in reality, parts are inter-dependent in a system and there are few systems in which the parts are independent from each other in all terms.

In this study, two issues of deteriorating parts and dependency of parts were investigated. A few works have been done on the optimization of deteriorating parts in single-component mode [1]. In these problems, parts are usually considered to have multiple modes or continuous deterioration functions [2]. One of the most popular continuous wear functions is gamma function. This function has been used to state deterioration of a single-component system or optimization of replacement time [3-8].

As noted above, most of the works done in this field are related to single-component systems and it is not easy to expand the results of these studies for multi-component systems. Maintenance policies for multi-component systems has been discussed in [9, 10]. In first article preventive maintenance optimization has been proposed in order to minimize system cost rate and in the second one grouping strategy maintenance is conducted and a case study of 16-component system has been illustrated.

Optimization in multi-component systems is much more complex than that in single-component ones. [11] is among the few studies on the optimization of replacement time when there is inter-dependency between parts. In this article, a series system with multi-mode parts was examined, which aimed to achieve optimal conditions for simultaneous replacement of parts. In addition, two-component series system was examined, which aimed to obtain the period of repair [12].

In terms of dependency, [13] article can be referred to. This paper examined the impact of correlation between parts on inspection duration of process. Since there are no data in this field due to its complexity, these studies have been far from practice. Also, none of the studies have addressed the following issue: How can deterioration function of parts be obtained in the mode of inter-dependency? Therefore, this study aimed to not only optimize the model, but also present a method for getting deterioration rates.

Consider a system with two parts like gear; at the end of each day, deterioration rate of its parts are examined and the decision about their replacement is made. In the method which is common in factories, when deterioration of each part reaches a certain extent (which is usually close to failure), that part is replaced; since deterioration rate is final, there is no failure during the use. It means that parts are replaced before any failure. Here, the difficulty is the ignorance of each part's deterioration on the deterioration of other parts.

This paper aimed to investigate the effect of dependency in deterioration rate on the optimal policy and present a method for optimizing replacement times of parts. To reach this goal, in Section 2, dynamic programming is used and optimal policy is determined. However, considering the point that the only data existing in the system are the time of deterioration limit of parts (replacement times of parts), in Section 3, a solution is presented for obtaining the deterioration rates. To get these failure rates, optimization-simulation was used in the present work so that a system with a discrete deterioration rate was simulated and deterioration rates were determined using simulated annealing (SA) algorithm in order to produce the same data as failure time. Two numerical examples are presented in section four and finally paper has been concluded in section five.

## 2. Investigating effect of dependency on optimal policy of replacement

### 2.1 Modeling

In this section, dynamic programming was used in order to, determining optimal policy, and investigating effects of dependency. Since programming horizon is infinite, two average reward and discount factor methods can be used; in this study, discount factor method was applied for programming.

$$u(i,j) = min \begin{cases} A(i,j) = \ \alpha \left( u\big(i + a(i,j), j + b(i,j)\big) \right) \\ B(i,j) = C_1 + A(1,j) \\ C(i,j) = C_2 + A(i,1) \\ D(i,j) = V + A(1,1) \end{cases} \quad (1)$$

$$u(i \geq L_1, j) = min \begin{cases} B(i,j) = C_1 + \alpha \left( u\big(1 + a(1,j), j + b(1,j)\big) \right) \\ D(i,j) = V + \alpha \left( u\big(1 + a(1,1), 1 + b(1,1)\big) \right) \end{cases} \quad (2)$$

where:

$u(i,j)$: $minimum\ cost\ of\ system\ when\ part\ 1\ is\ in\ state\ i\ and\ part\ 2\ is\ in\ state\ j$

$i$: $Deterioration\ rate\ of\ part\ 1$

$j$: $Deterioration\ rate\ of\ part\ 2$

$L_1$: $Replacement\ limit\ of\ part\ 1$

$C_k$: $Replacement\ cost\ of\ part\ k$

$V$: $Simultanious\ \ replacement\ cost$

$\alpha$: $Discount\ coeficient$

$A(i,j)$: $Proceeding\ cost$

$B(i,j)$: $Cost\ of\ replacment\ selection\ of\ part\ 1$

$C(i,j)$: $Cost\ of\ replacement\ selection\ of\ part\ 2$

$D(i,j)$: $Cost\ of\ replacment\ selection\ of\ both\ parts$

## 2.2 Determining optimal policy

In order to obtain the optimal policy, first, it has been proved by using induction that the ratio of proceeding cost considering deterioration rate of parts is ascending. Then, considering the constant ratio of cost of part replacement to its deterioration rate, the optimal policy is obtained. The proof given in the appendix A results in obtaining the following facts:

- Selection cost A (proceeding without replacement) is ascending in proportion to $i, j$.
- Selection cost B (replacement of part 1) is constant relative to $i$ and ascending to $j$.
- Selection cost C (replacement of part 2) is ascending relative to $i$ and constant to $j$.
- Selection cost D (replacement of both parts) is constant to $i, j$.

According to the ascending cost of "proceeding without replacement" and constant cost of replacement of a part in proportion to its deterioration rate, it can be proved that, per deterioration rate of part 2, there is a limit $i_j^*$ for the deterioration of part 1; from then on, "proceeding without replacement" is not optimal. On the other hand, considering the constant cost of "replacement of part 1" relative to its deterioration rate, if selecting "replacement of part 1" is optimal at point $(i, j)$, "replacement of part 1" will be optimal for points $(i' > i, j' = j)$. The mentioned reasoning also holds for selecting "replacement of part 2".

Selection cost of "simultaneous replacement of both parts" relative to their deterioration rate is constant. Therefore, given that the other three selections are ascending or constant relative to the deterioration rate of parts, if "replacement of both parts" is an optimal selection at one point, "simultaneous replacement of both parts" will be optimal per increase in the deterioration of parts.

In summary, optimal policy is as follows: per deterioration rate of part 2, there is $i_j^*$ limit for deterioration of part 1, at which "replacement of part 1" or "both parts" is optimal and, with increasing deterioration rate of part 1, the selection is still optimal. Moreover, per increase in the deterioration of part 1, there is $j_i^*$ limit for the deterioration of part 2, at which "replacement of part 2" or "both parts" is optimal and, with increasing deterioration rate of part 2, the selection is still optimal and its replacement is optimal $(i, 1)$ at point $(i, j)$. Also, it can be proved that, if replacement part 2 is optimal at point $(i, j)$, then proceeding without replacement will be optimal at point $(i, 1)$, because if replacement of part 2 is optimal at point $(i, j)$, the system will go to mode $(i, 1)$. Thus, replacement of part 1 cannot be optimal at this point; if this replacement is optimal, replacement of both will occur at point $(i, j)$.

To solve the model, the following linear programming can be used:

$$Max\ z = \sum \sum u(i, j) \quad (3)$$

$S.T.$

$$u(i, j) \leq \alpha \left( u\left( i + a(i, j), j + b(i, j) \right) \right) \quad (4)$$

$$u(i, j) \leq C_1 + \alpha \left( u\left( 1 + a(1, j), j + b(1, j) \right) \right) \quad (5)$$

$$u(i, j) \leq C_2 + \alpha \left( u\left( i + a(i, 1), 1 + b(i, 1) \right) \right) \quad (6)$$

$$u(i, j) \leq V + \alpha \left( u\left( 1 + a(1,1), 1 + b(1,1) \right) \right) \quad (7)$$

Using the values obtained by solving linear programming for $u(i, j)$, optimal limits of replacement can be obtained. To do this, the following equations should be used; i.e. cost of proceeding and replacing part 1 is compared and $i'_j$ is obtained. Then, cost of proceeding and simultaneously replacing both parts is compared and $i''_j$ is calculated. Minimum values of them is replacement limit $i^*_j$. If $i^*_j$ is minimum, replacement of part 1 is optimal and, if $i''_j$ is minimum, simultaneous replacement of both parts is optimal.

$$i'_j \in \{ i | C_1 + u(1, j) = u(i, j) \} \quad (8)$$

$$i''_j \in \{ i | V + u(1,1) = u(i, j) \} \quad (9)$$

$$i^*_j = Min\{ i'_j, i''_j \} \quad (10)$$

The same should be followed for obtaining replacement limits of part 2.

## 3. Obtaining deterioration rates using previous failure data

Here our aim is to evaluate failure rate modes of system parts by employing replacement data available from the past. A proposed model is presented in next subsection and then the selection criteria for appropriate algorithm is discussed.

### 3.1 Modeling

As mentioned in the introduction, the only data which are available in a system are replacement time of parts in the past with replacement policy when approaching the failure limit. Besides, the problem nature is continuous. But, for optimization using dynamic programming deterioration rate of parts in several modes is required.

To obtain deterioration rates, first, several modes are considered for each part. Then, using mathematical programming, deterioration rates are calculated so that the sum of squared deviation from the past failure data would be minimum. Mathematical modeling of this problem is very complex and so, the estimated model of the problem is as follows:

$N = number\ of\ avaiable\ past\ data\ points$

$N_i = number\ of\ part\ i's\ \ failure\ times$

$I_{i,n} =\ nth\ failure\ period\ of\ part\ i$

$m_i =\ number\ of\ working\ modes\ of\ part\ i$

$L_i = replacement\ limit\ of\ part\ i$

$a_{ij} = deterioration\ rate\ of\ part\ 1\ when\ part\ 1\ is\ in\ \ state\ i\ and\ part\ 2\ is\ in\ state\ j$

$b_{ij} = deterioration\ rate\ of\ part\ 2\ when\ part\ 1\ is\ in\ \ state\ i\ and\ part\ 2\ is\ in\ state\ j$

$x_{nij} = duration\ time\ in\ nth\ period\ in\ which\ part\ 1\ is\ in\ \ state\ i\ and\ part\ 2\ is\ in\ state\ j$

$L_{1i} = mode\ ith's\ deterioration\ rate\ of\ part\ 1$

$L_{2j} = mode\ jth's\ deterioration\ rate\ of\ part\ 2$

$T_{in} = nth\ replacement\ time\ of\ part\ i$

$\varepsilon_n = the\ model\ data\ deviation\ from\ historical\ data$

$$Min\ V = \sum_{i=1}^{2}\sum_{n=1}^{N_i} \varepsilon_{in}^2 \tag{11}$$

$S.T$

$$\sum_{k=I_{1,t-1}+1}^{I_{1,t}} \sum_{j=1}^{m_2}\sum_{i=1}^{m_1} a_{ij}.x_{kij} \geq L_1 \qquad ,t = 1,2,\dots,N_1 \tag{12}$$

$$\sum_{k=I_{1,t-1}+1}^{I_{1,t}} \sum_{j=1}^{m_2}\sum_{i=1}^{m_1} a_{ij}.x_{kij} \leq L_1 + A_{1t} \qquad ,t = 1,2,\dots,N_1 \tag{13}$$

$$A_{1t} = \underset{i,j}{Max}\left\{a_{ij}.Min\{1, x_{tij}\}\right\}$$

$$\sum_{k=I_{2,t-1}+1}^{I_{2,t}} \sum_{j=1}^{m_2}\sum_{i=1}^{m_1} b_{ij}.x_{kij} \geq L_2 \qquad ,t = 1,2,\dots,N_2 \tag{14}$$

$$\sum_{k=I_{2,t-1}+1}^{I_{2,t}} \sum_{j=1}^{m_2}\sum_{i=1}^{m_1} b_{ij}.x_{kij} \leq L_2 + A_{2t} \qquad ,t = 1,2,\dots,N_2 \tag{15}$$

$$A_{2t} = \underset{i,j}{Max}\left\{b_{ij}.Min\{1, x_{tij}\}\right\}$$

$$x_{n,ij} \leq M.\left(x_{n,i-1j} + x_{n,ij-1} + x_{n,i-1j-1}\right) \qquad ,n = 1,2,\dots,N\ ,i = 2\dots m_1\ ,j = 2\dots m_2 \tag{16}$$

$$x_{n,1j} \leq M.\left(x_{n,1j-1} + x_{n-1,m_1j} + x_{n-1,m_1j-1}\right) \qquad ,n = 1,2,\dots,N\ ,j = 2\dots m_2 \tag{17}$$

$$x_{n,i1} \leq M.\left(x_{n,i-11} + x_{n-1,im_2} + x_{n-1,i-1m_2}\right) \qquad ,n = 1,2,\dots,N\ ,i = 2\dots m_1 \tag{18}$$

$$x_{n,11} \leq M.\left(x_{n-1,m_11} + x_{n-1,1m_2} + x_{n-1,m_1m_2}\right) \qquad ,n = 2,\dots,N \tag{19}$$

$$x_{n,i+1j}.x_{n,ij+1} = 0 \qquad ,i = 1 \dots m_1 - 1\,, j = 1 \dots m_2 - 1 \tag{20}$$

$$M.Max\left\{\sum_{k=I_{1,t}+1}^{n-1}\sum_{j=1}^{m_2}\sum_{i=1}^{m_1} a_{ij}.x_{kij} + \sum_{j=1}^{j'}\sum_{i=1}^{i'-1} a_{ij}.x_{nij} - L_{1i'-1}, 0\right\} \geq x_{ni'j'} \tag{21}$$
$$,n = 1 \dots N, i' = 1 \dots m_1, j' = 1 \dots m_2, t = Max\{k: I_{1k} < n\}$$

$$M.Max\left\{L_{1i'} - \sum_{k=I_{1,t}+1}^{n-1}\sum_{j=1}^{m_2}\sum_{i=1}^{m_1} a_{ij}.x_{kij} + \sum_{j=1}^{j'}\sum_{i=1}^{i'-1} a_{ij}.x_{nij}\,, 0\right\} \geq x_{ni'j'} \tag{22}$$
$$,n = 1 \dots N\,, i' = 1 \dots m_1 \;\;, j' = 1 \dots m_2\,, t = Max\{k: I_{1k} < n\}$$

$$M.Max\left\{\sum_{k=I_{2,t}+1}^{n-1}\sum_{j=1}^{m_2}\sum_{i=1}^{m_1} b_{ij}.x_{kij} + \sum_{j=1}^{j'-1}\sum_{i=1}^{i'} b_{ij}.x_{nij} - L_{2j'-1}, 0\right\} \geq x_{ni'j'} \tag{23}$$
$$,n = 1 \dots N\,, i' = 1 \dots m_1 \;\;, j' = 1 \dots m_2\,, t = Max\{k: I_{2k} < n\}$$

$$M.Max\left\{L_{2j'} - \sum_{k=I_{2,t}+1}^{n-1}\sum_{j=1}^{m_2}\sum_{i=1}^{m_1} b_{ij}.x_{kij} + \sum_{j=1}^{j'-1}\sum_{i=1}^{i'} b_{ij}.x_{nij}\,, 0\right\} \geq x_{ni'j'} \tag{24}$$
$$,n = 1 \dots N\,, i' = 1 \dots m_1 \;\;, j' = 1 \dots m_2\,, t = Max\{k: I_{2k} < n\}$$

$$\sum_{k=I_{i,n-1}+1}^{I_{i,n}}\sum_{j=1}^{m_2}\sum_{i=1}^{m_1} x_{kij} + \varepsilon_{in} = T_{in} - T_{i,n-1} \qquad ,n = 1 \dots N_i \tag{25}$$

$a_{ij}, b_{ij} > 0\,, Integer \;\; ; \;\; x_{n,i,j} \geq 0\,, Integer$

$x_{n,i,j} = 0\,, n = 0 \vee i = 0 \vee j = 0 \;\; ; \;\; M \text{ represents some very large number}$

In this model, Equation (11) is the objective function, which indicates the sum of squared deviations from the failure time obtained from the model and failure time of previous data. Constraints (12), (13), (14), and (15) indicate that, at failure time, deterioration of parts should exceed their failure limits. Constraints (16), (17), (18), and (19) show that each part can only go to the next mode and mode decrease or increase by more than one is not allowed. Constraints (21), (22), (23), and (24) are added to the model to only enable $x_{ni'j'}$ to take a number when the system is in mode $(i', j')$; otherwise, it will be 0.

Finally, Constraint (25) indicates the deviation amount of historical data from the model data. This model is non-linear and integer; so, common methods cannot be used for solving it. Hence, "optimization-simulation" method was used in this paper. One of the advantages of this method is that it does not need complex and insoluble modeling.

In this paper, the combination of simulation and meta-heuristic algorithms was used for optimization. Thus, values of decision variables were generated by meta-heuristic algorithm and the objective function was calculated by simulation. The obtained objective function was again investigated by meta-heuristic algorithm and new values were generated for the variables. This cycle continued until reaching the stopping condition of meta-heuristic algorithm. Comparison of meta-heuristic algorithms used in preventive maintenance milieu can be found in [14].

Simulated algorithm of this model is given in Figure 1. Output of this flowchart considering the values of deterioration rate as input shows time of reaching failure limit and type of the part reaching this limit (similar to the available failure data from the past). Then, by calculating the squared deviation of the time obtained from the present times using the past data, the value of objective function is calculated and used in meta-heuristic algorithm.

### 3.2 Selecting appropriate meta-heuristic algorithm

It has been proved that, under certain assumptions, there is no algorithm which is superior to others in terms of all optimization problems. One of the most important aspects of designing meta-heuristic algorithms is the analysis of response space of the optimization problem, because effect of a meta-heuristic algorithm depends on the characteristics of response space such as roughness, fluctuation, convexity, etc. Analysis of response space is done to predict the behavior of the search components of a meta-heuristic algorithm including search operator, representatives, and objective function. So, this analysis helps in the better selection of objective function, representatives, and search operator [15, 16]. There are different indicators for analyzing response space; here, three indicators are used:

Distribution in the objective space: To investigate distribution in the objective space, amplitude index is used. This index is a set of equal desired responses with relative difference between the best and worst objective functions from the set, which is defined as follows:

$$Amp(P) = \frac{\left(|P|.\left(\max f(s) - min f(s)\right)\right)}{\sum f(s)} \qquad (26)$$

where p is size of population and $f(s)$ is value of obtained objective function. The lower this index, the easier to solve this problem using meta-heuristic algorithm based on a response [15].

Length of the walks: Length of walks equals the number of walks passed from the initial response to the local optimal response. The more the value of this parameter, the less the fluctuation and roughness of the response space would be. In a high fluctuation and rough space, the number of local optimal points is high and therefore mean length of walks is less. If local optimal points with the policy of a change per iteration (e.g. SA algorithm) are obtained, then, the higher this value, the better the movement in the objective space with one change [15].

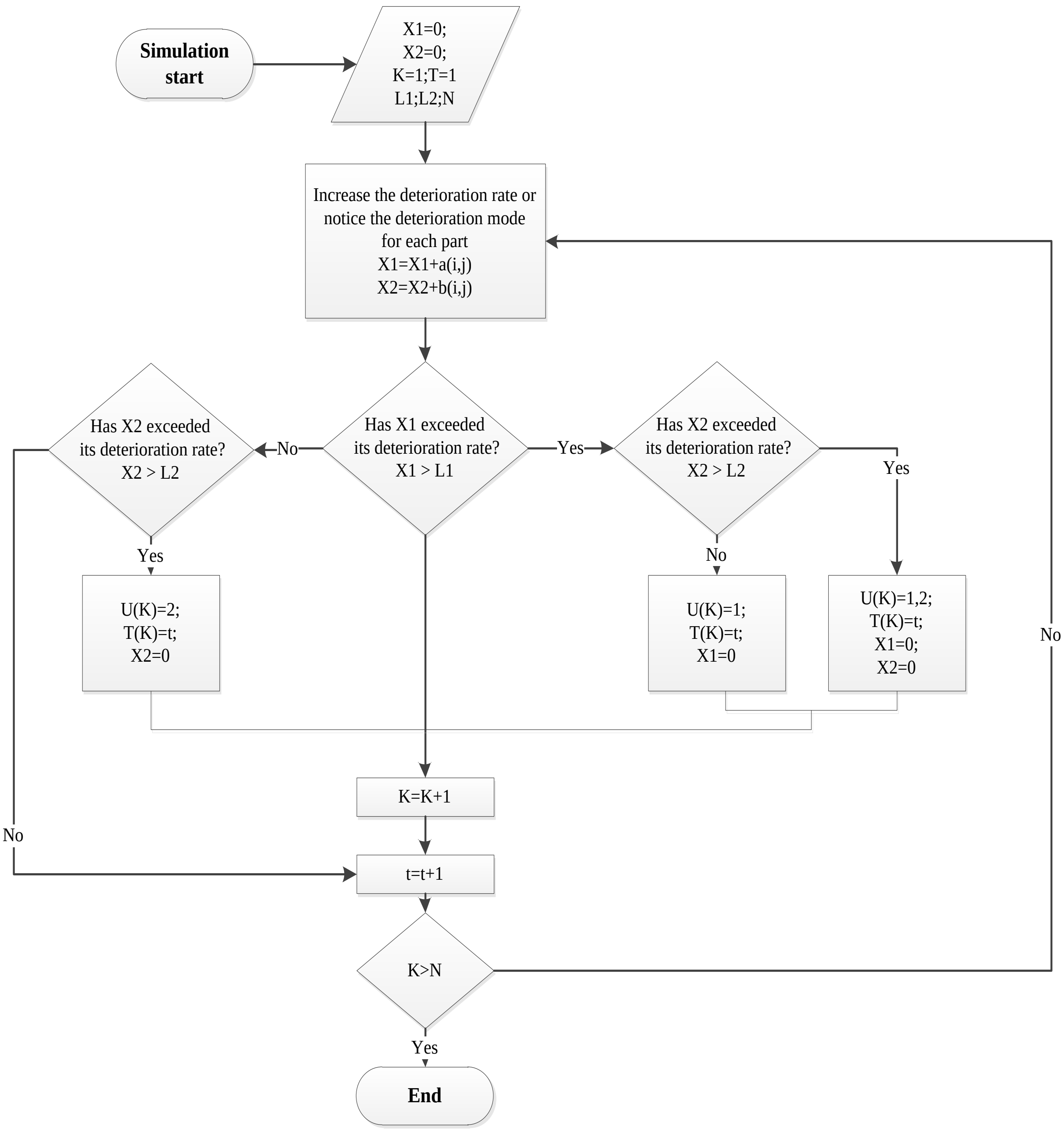

*Figure 1 Simulation process*

Autocorrelation function: One of the major and effective factors for local search process is roughness value of response space. The index which is used to measure roughness rate of the objective space is called autocorrelation function. If there is no correlation between distance of responses and value of objective function, the response space is rough. Search in such a space using local search methods is very difficult. In contrast, the smooth response space is very convenient for searching in the neighborhood.

Therefore, value of autocorrelation helps decide on search in the neighborhood. In a rough response space, there is no correlation between values of objective function of two neighboring points and it cannot help guide the search process. In contrast, in a smooth response space, there is correlation between objective function values of two neighboring points and this correlation helps the process to move toward the optimal response.

In order to calculate autocorrelation of response space, the whole response space of an example is required to be investigated, which is not possible for large sizes.[16]. So, Weinberger [17] suggested random walking method for estimating autocorrelation. This method starts with an optional initial response and a response is randomly generated in its neighborhood. This process is repeated for $m$ repetitions until the objective function value is calculated. By obtaining values of the objective function, the autocorrelation value can be calculated using Equation below:

$$r(1) \approx \frac{1}{\sigma^2(m-1)} \sum_{t=1}^{m-1} \left(f(x_t) - \bar{f}\right)\left(f(x_{t+1}) - \bar{f}\right) \tag{27}$$

According to the indices of meta-heuristic algorithms discussed here, SA algorithm has been employed to obtain simulated data. The parameters of SA algorithm used in this study are given below.

**Objective function:** Sum of squared difference of simulated and real failure time from the past.

**Initial temperature:** Here, initial temperature is obtained as follows: first, the sufficient number of initial population and a point are generated in its neighborhood. Then, the following equation is used to obtain the initial temperature, in which $\Delta^{+}$ equals mean increase in the objective function of the neighborhood, $\mathrm{m}_1$ is the number of responses in which the value of objective function is reduced in the neighborhood, $\mathrm{m}_2$ is the number of responses in which the objective function values are increased in the neighborhood, and $\mathrm{a}_0$ equals the percentage of the population which should be accepted at the initial temperature. This value must be selected between 0.4 and 0.5, which was 0.5 in this study.

$$T_0 = \frac{\Delta^{+}}{\ln\left(\frac{m_1(a_0 - 1)}{m_2} + a_0\right)} \tag{28}$$

**Geometric decrease temperature function:** $T(t+1) = \alpha T(t)$ (29)

**Number of iterations**: Number of repetitions at each temperature is constant and equals n.

**Stopping condition: Reaching n iterations**

Below, numerical examples are presented and the results are discussed.

## 4. Numerical example

In this section, two examples are given to examine the efficiency of the method.

### Example 1

Assume that data of failure time of the parts in a two-component system (such as two involved gears) are as in Table 1 and failure limit of each part is 0.9 mm. Other data are as follows:

Replacement cost of part 1: 100
Replacement cost of part 2: 120
Simultaneous replacement cost of both parts: 220

So, deterioration rates must be first calculated using data in a discrete way (for solving by dynamic programming, discrete forms of these rates are needed). To calculate these rates, as stated above, optimization-simulation can be used. For this example, the value of $Amp(P)$ index was 3.05 according to the eq.(26) and length of the walks equivalent to 8.2 walks. Moreover, the obtained autocorrelation function for this issue is 0.93, which is high and properly shows that the response space is smooth and low-fluctuation; therefore, the algorithm based on a local change and search like SA can be easily accepted.

To obtain the initial temperature of SA algorithm for this example, the given relationship was employed and 220° was achieved with 1000 repetitions. Also, sensitivity analysis was performed on other parameters. Results of solving this example by annealing algorithm for obtaining deterioration rates in each mode considering 10 modes for each part are given in the following table. In this solution, three rates of temperature reduction as 0.98, 0.99, and 0.999 and three repetitions at each temperature as 15, 20, and 25 were implemented in every mode of 10 times of the algorithm. Here, in order to examine the influence of parameters on the optimal response, stopping condition with 30,000 repetitions was considered.

*Table1 Failure data for example 1*

| part | Failure time | Part | Failure time | Part | Failure time | part | Failure time |
|---|---|---|---|---|---|---|---|
| 1 | 26 | 2 | 154 | 1 | 303 | 2 | 432 |
| 2 | 28 | 1 | 172 | 2 | 306 | 1 | 447 |
| 1 | 46 | 2 | 175 | 1 | 321 | 2 | 451 |
| 2 | 49 | 1 | 190 | 2 | 325 | 1 | 464 |
| 1 | 64 | 2 | 193 | 1 | 338 | 2 | 468 |
| 2 | 68 | 1 | 209 | 2 | 342 | 1 | 481 |
| 1 | 81 | 2 | 211 | 1 | 355 | 2 | 484 |
| 2 | 84 | 1 | 229 | 2 | 358 | 1 | 499 |
| 1 | 99 | 2 | 232 | 1 | 374 | 2 | 503 |
| 2 | 103 | 1 | 247 | 2 | 376 | 1 | 516 |
| 1 | 116 | 2 | 251 | 1 | 394 | 2 | 520 |
| 2 | 120 | 1 | 264 | 2 | 397 | 1 | 533 |
| 1 | 133 | 2 | 267 | 1 | 412 | 2 | 537 |
| 2 | 136 | 1 | 283 | 2 | 416 | 1 | 550 |

As can be seen in Table 2, the best result (minimum sum of squared errors) was related to the temperature reduction coefficient of 0.999 and number of repetition at each temperature of 20. The worst results were related to temperature reduction coefficient of 0.99 and number of repetitions at each temperature of 10. This issue can be due to rapid temperature decrease, which results in the improper search of the space by the algorithm; but, the results are close to each other and are not greatly different.

*Table 2 Results of SA algorithm*

| | | | Results of algorithm | | | | | | | | | | |
|---|---|---|---|---|---|---|---|---|---|---|---|---|---|
| $n$ | $T_0$ | $\alpha$ | 1 | 2 | 3 | 4 | 5 | 6 | 7 | 8 | 9 | 10 | Average |
| 10 | 220 | 0.98 | 112 | 120 | 124 | 104 | 115 | 95 | 135 | 103 | 103 | 103 | 111.4 |
| 10 | 220 | 0.99 | 144 | 86 | 138 | 106 | 103 | 144 | 117 | 99 | 104 | 147 | 118.8 |
| 10 | 220 | 0.999 | 103 | 114 | 115 | 127 | 98 | 131 | 115 | 115 | 115 | 106 | 113.9 |
| 15 | 220 | 0.98 | 143 | 103 | 108 | 112 | 106 | 106 | 103 | 103 | 95 | 103 | 108.2 |
| 15 | 220 | 0.99 | 102 | 114 | 103 | 104 | 103 | 128 | 103 | 112 | 104 | 102 | 107.5 |
| 15 | 220 | 0.999 | 103 | 103 | 104 | 104 | 118 | 104 | 98 | 104 | 103 | 104 | 104.5 |
| 20 | 220 | 0.98 | 120 | 104 | 135 | 103 | 115 | 104 | 103 | 115 | 105 | 115 | 111.9 |
| 20 | 220 | 0.99 | 100 | 97 | 102 | 99 | 176 | 98 | 119 | 123 | 110 | 109 | 113.3 |
| 20 | 220 | 0.999 | 101 | 101 | 114 | 104 | 104 | 101 | 104 | 104 | 103 | 103 | 103.9 |

Below, to find the optimal strategy, one of the solutions obtained by annealing algorithm was used, in which the sum of squared error was equal to 101. This response is given in Tables 3 and 4. Using the deterioration rates in Tables 3 and 4, the dynamic programming whose solution and modeling were discussed in the section 2, with discount factor of 0.95, the example was resolved.

*Table 3 Deterioration rate of part 1*

| $a(i,j)$ | | Deterioration of part 2 (0.01mm) | | | | | | | | | |
|---|---|---|---|---|---|---|---|---|---|---|---|
| | | 1-9 | 10-18 | 19-27 | 28-36 | 37-45 | 45-54 | 55-64 | 65-72 | 73-81 | 82-90 |
| Deterioration of part 1 (0.01mm) | 1-9 | 1 | 1 | 3 | 3 | 4 | 4 | 4 | 5 | 6 | 7 |
| | 10-18 | 1 | 1 | 3 | 3 | 4 | 4 | 5 | 5 | 8 | 12 |
| | 19-27 | 3 | 3 | 3 | 3 | 4 | 5 | 6 | 8 | 9 | 12 |
| | 28-36 | 3 | 3 | 3 | 4 | 4 | 5 | 6 | 9 | 9 | 12 |
| | 37-45 | 3 | 3 | 4 | 4 | 5 | 6 | 7 | 10 | 11 | 12 |
| | 46-54 | 4 | 7 | 8 | 8 | 8 | 9 | 9 | 11 | 11 | 12 |
| | 55-64 | 4 | 9 | 10 | 11 | 11 | 11 | 11 | 11 | 12 | 13 |
| | 65-72 | 6 | 11 | 12 | 13 | 13 | 13 | 13 | 13 | 13 | 14 |
| | 73-81 | 7 | 12 | 12 | 13 | 14 | 14 | 14 | 14 | 14 | 14 |
| | 82-90 | 7 | 12 | 12 | 14 | 14 | 14 | 14 | 14 | 14 | 14 |

*Table 4 Deterioration rate of part 2*

| $b(i,j)$ | | Deterioration of part 2 (0.01mm) | | | | | | | | | |
|---|---|---|---|---|---|---|---|---|---|---|---|
| | | 1-9 | 10-18 | 19-27 | 28-36 | 37-45 | 46-54 | 55-64 | 65-72 | 73-81 | 82-90 |
| Deterioration of part 1 (0.01mm) | 1-9 | 1 | 1 | 1 | 1 | 4 | 7 | 7 | 9 | 9 | 9 |
| | 10-18 | 1 | 2 | 2 | 2 | 4 | 7 | 7 | 9 | 10 | 11 |
| | 19-27 | 1 | 2 | 4 | 4 | 4 | 8 | 8 | 9 | 11 | 12 |
| | 28-36 | 4 | 5 | 5 | 5 | 6 | 9 | 11 | 12 | 13 | 13 |
| | 37-45 | 5 | 6 | 7 | 7 | 8 | 9 | 11 | 13 | 13 | 13 |
| | 46-54 | 6 | 7 | 8 | 9 | 10 | 10 | 11 | 13 | 13 | 13 |
| | 55-64 | 6 | 8 | 11 | 11 | 12 | 13 | 13 | 13 | 13 | 13 |
| | 65-72 | 6 | 8 | 11 | 11 | 12 | 13 | 13 | 13 | 13 | 13 |
| | 73-81 | 6 | 8 | 13 | 13 | 13 | 13 | 13 | 13 | 13 | 13 |
| | 82-90 | 7 | 9 | 13 | 13 | 13 | 13 | 13 | 14 | 14 | 14 |

The obtained optimal policy is shown in Figure 2. In this figure, blue shows proceeding without replacement, red is replacement of part 1, green is replacement of part 2, and purple is simultaneous replacement of both parts. As demonstrated in the figure, there are some areas, in which the optimal decision is to replace one of the parts although both parts are sound, can still continue to work, and may not fail during working. The reason is the dependency between the parts, since if the deteriorated part is not replaced, deteriorating rate of the other part will increase.

As can be seen in Figure 2 and was already proved, by increasing the deterioration rate of each part, there will be a limit for the deterioration of part one and it should be replaced after that limit. Also, at any point at which simultaneous replacement of parts is optimal, with increasing deterioration rate of each part, simultaneous replacement of parts is still optimal.

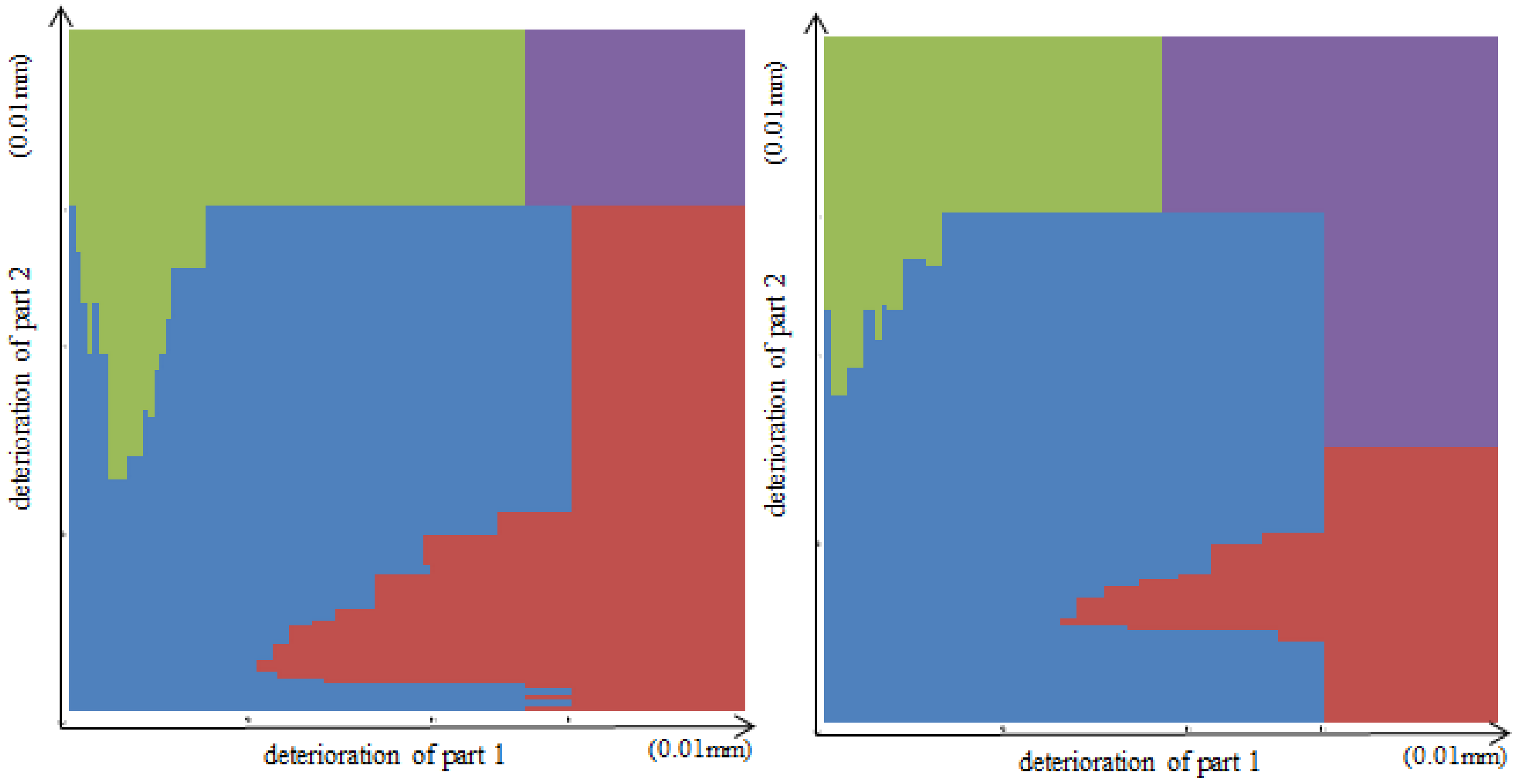

(a) (b)

*Figure2 Optimal policy for example one: (a) without economic dependency (b) with economic dependency*

In practice, replacing system parts are economically interdependent which means solo replacement increases process downtime. In multi-component systems one may replaces some parts simultaneously in order to avoid extra setup cost [18]. In this example, if economic dependency was also added so that simultaneous replacement cost of both parts became 200, instead of 220, the optimal policy would be as in Figure 2(b). By comparing figures 2(a) and 2(b), we can conclude that the general policy is almost the same. Moreover, it can be seen that joint replacement area under economic dependency circumstances has increased. This was expected because economic interdependency results in more simultaneous replacements.

Mean cost of part replacement which can be calculated from the historical data given in Table 1 is equal to 11.874, while replacement cost using the obtained strategy is equal to 9.525; it means 20% cost reduction.

**Example 2**

Consider another example, in which failure time data of parts are as in Table 5 and failure limit of each part is equal to 0.9 mm. Other details are as follows:

Replacement cost of part 1: 100
Replacement cost of part 2: 300
Simultaneous replacement cost of both parts: 400

Here, similar to the previous example, first, deterioration rate should be obtained from the failure data in the past tabulating in table 5. To do this, as described previously, SA algorithm was used. The obtained deterioration rates using this method are given in Tables 6 and 7 (for these rates, the objective function value was equal to 1).

*Table 5 Failure data for example 2*

| part | Failure time | Part | Failure time |
|---|---|---|---|
| 1 | 21 | 2 | 68 |
| 2 | 25 | 1 | 74 |
| 1 | 35 | 2 | 82 |
| 2 | 40 | 1 | 86 |
| 1 | 48 | 2 | 97 |
| 2 | 54 | 1 | 99 |
| 1 | 61 | 1,2 | 114 |

*Table 6 Deterioration rate of part 1*

| $a(i,j)$ | | *Deterioration of part 2 (0.01mm)* | | | | | | | | | |
|---|---|---|---|---|---|---|---|---|---|---|---|
| | | 1-9 | 10-18 | 19-27 | 28-36 | 37-45 | 45-54 | 55-64 | 65-72 | 73-81 | 82-90 |
| *Deterioration of part 1 (0.01mm)* | 1-9 | 1 | 1 | 1 | 3 | 5 | 5 | 8 | 8 | 8 | 10 |
| | 10-18 | 2 | 3 | 5 | 5 | 6 | 7 | 8 | 8 | 8 | 10 |
| | 19-27 | 2 | 4 | 5 | 6 | 6 | 7 | 9 | 10 | 11 | 11 |
| | 28-36 | 2 | 4 | 5 | 7 | 9 | 9 | 10 | 10 | 11 | 11 |
| | 37-45 | 2 | 7 | 7 | 7 | 10 | 10 | 10 | 10 | 12 | 12 |
| | 46-54 | 2 | 7 | 7 | 8 | 10 | 11 | 11 | 11 | 12 | 13 |
| | 55-64 | 5 | 7 | 7 | 10 | 10 | 13 | 13 | 13 | 13 | 13 |
| | 65-72 | 9 | 10 | 12 | 12 | 12 | 13 | 14 | 14 | 14 | 14 |
| | 73-81 | 9 | 12 | 12 | 12 | 12 | 13 | 14 | 14 | 14 | 14 |
| | 82-90 | 12 | 12 | 12 | 13 | 13 | 14 | 14 | 14 | 14 | 14 |

*Table 7 Deterioration rate of part 2*

| $b(i,j)$ | | Deterioration of part 2 (0.01mm) | | | | | | | | | |
|---|---|---|---|---|---|---|---|---|---|---|---|
| | | 1-9 | 10-18 | 19-27 | 28-36 | 37-45 | 45-54 | 55-64 | 65-72 | 73-81 | 82-90 |
| Deterioration of part 1 (0.01mm) | 1-9 | 1 | 1 | 1 | 1 | 1 | 4 | 6 | 7 | 8 | 10 |
| | 10-18 | 1 | 2 | 2 | 3 | 7 | 8 | 8 | 8 | 8 | 13 |
| | 19-27 | 1 | 2 | 2 | 3 | 7 | 9 | 9 | 10 | 12 | 13 |
| | 28-36 | 3 | 3 | 3 | 4 | 7 | 12 | 12 | 12 | 13 | 13 |
| | 37-45 | 3 | 3 | 4 | 4 | 8 | 12 | 12 | 12 | 13 | 13 |
| | 46-54 | 3 | 3 | 4 | 7 | 8 | 12 | 12 | 12 | 13 | 13 |
| | 55-64 | 3 | 4 | 6 | 8 | 8 | 12 | 12 | 12 | 14 | 14 |
| | 65-72 | 3 | 7 | 10 | 11 | 12 | 12 | 12 | 12 | 14 | 14 |
| | 73-81 | 9 | 10 | 10 | 11 | 12 | 12 | 13 | 13 | 14 | 14 |
| | 82-90 | 10 | 10 | 10 | 12 | 12 | 12 | 13 | 14 | 14 | 14 |

Now, like the previous example, replacement limits of each part were obtained using dynamic programming. In this example, discount factor was considered 0.95. Optimal replacement limits are shown in Figure 3. As can be seen in this figure similar to the previous example, there were areas in which replacement should be done although both parts were sound. Because if no replacement occurred, deterioration rate of both parts would increase due to the deformation in the part and consequently the system cost would increase.

In this example, per deterioration rate in part 2, there was a limit for the deteriorating of part 1 and part 1 should be replaced after that limit. Also, at any point at which simultaneous replacement of parts was optimal, with increasing the deterioration rate of each part, simultaneous replacement of the parts was still optimal.

Mean cost of parts replacement which can be calculated from the historical data given in Table 5 was equal to 25.44, while mean cost of replacement using the given strategy was equal to 16.7; i.e. 34% cost reduction. The reason of this increase in cost reduction relative to the previous example was high cost difference between replacing parts 1 and 2, which led to preventive replacements (part 1 was replaced sooner, because it was less costly).

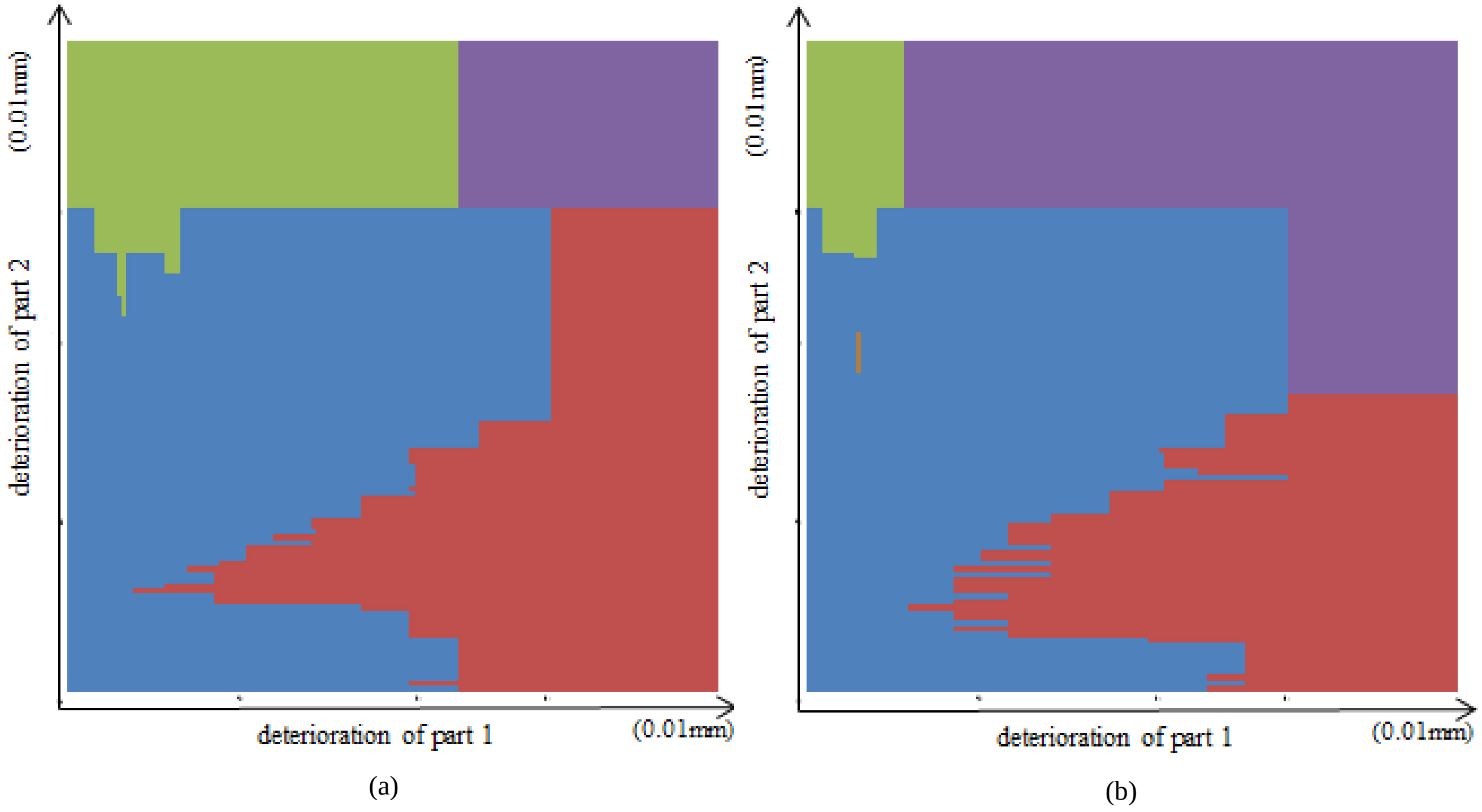

*Figure3 Optimal policy for example two: (a) without economic dependency (b) with economic dependency*

In this example, if economic dependency was also added so that simultaneous replacement cost of both parts became 350, instead of 400, the optimal policy would be as in Figure 3(b). As can be seen in this figure, the general policy was almost the same; but, the area related to simultaneous replacement was increased. Mean cost of using the current policy using the past data was equal to 25, while using the obtained policy, the mean cost became 15.66, which represented 37% cost reduction.

## 5. Conclusion

In this study, optimal replacement time of deteriorating parts which depends on each other and has a definite deterioration rate was examined and attempts were made to present a practical method for solving this issue. Dynamic programming was used for modeling and optimizing the problem and the optimal policy was obtained. However, in reality, deterioration rates are not available, deterioration is continuous, and the only available information of the system is replacement time of parts in the past. To solve this problem, a system with a discrete deterioration rate was stimulated; in this simulation according to the input rates, times were obtained for the failure of parts.

In order to obtain a system close to the real problem, SA algorithm was used so that the objective function was the deviation rate and calculation of the objective was done by stimulation. Finally, two examples were presented and their solution approved our analysis about optimal policy. Moreover, it was demonstrated that, in the case of using the presented method, cost would be considerably reduced. So we suggest in systems with two or more dependent parts, using SA algorithm in order to obtain simulated failure time points can be costly beneficial.

## Appendix A: Proving ascending nature of cost relative to deterioration rate

For the purpose of proving, induction was used.

According to the definition of the model, it is clear that $\mathrm{u}(\mathrm{i},\mathrm{j}) \leq \mathrm{u}(\mathrm{L}_1,\mathrm{j})$; therefore, $\mathrm{u}(\mathrm{L}_1-1,\mathrm{j}) \leq \mathrm{u}(\mathrm{L}_1,\mathrm{j})$.

Premise:

$$\forall\, i \geq k+1\,, j\ :\ u(i,j) \leq u(i+1,j) \qquad (A1)$$

Proposition:

$$u(k,j) \leq u(k+1,j) \qquad (A2)$$

The following is done for proving:

$$A(k+1,j) - A(k,j) = \alpha\left(u\big(k+1+a(i,j), j+b(i,j)\big) - u\big(k+a(i,j), j+b(i,j)\big)\right) \geq 0 \qquad (A3)$$

Since $a(i,j) \geq 1$, according to the induction premise, the expression on the right side is larger than zero.

$$B(k+1,j) - B(k,j) = C_1 + A(1,j) - C_1 - A(1,j) = 0 \qquad (A5)$$

$$C(k+1,j) - C(k,j) = C_2 + A(k+1,1) - C_2 - A(k,1) = A(k+1,1) - A(k,1) \geq 0 \qquad (A6)$$

$$D(k+1,j) - D(k,j) = V + A(1,1) - V - A(1,1) = 0 \qquad (A7)$$

On the other hand:

$$u(k,j) = min\begin{cases} A(k,j) \\ B(k,j) \\ C(k,j) \\ D(k,j) \end{cases} \quad \text{و} \quad u(k+1,j)$$

$$= min\begin{cases} A(k+1,j) \\ B(k+1,j) \\ C(k+1,j) \\ D(k+1,j) \end{cases} \qquad (A8)$$

So, $u(k, j) \leq u(k+1, j)$. Thus, it is proved that function $u(i, j)$ is ascending relative to the deterioration rate of parts. Ascending nature of function $u(i, j)$ results in the ascending cost of selecting "to proceed without replacement $A(i, j)$".